\documentclass[aps,twocolumn,showpacs,preprintnumbers,prb]{revtex4}

\usepackage{graphicx}
\usepackage{dcolumn}
\usepackage{bm}

\begin{document}

\title{Lifetimes of Shockley electrons and holes at the Cu(111)
surface}
\author{M.~G.~Vergniory$^1$, J.~M.~Pitarke$^{1,2}$, and
S.~Crampin$^3$} 
\affiliation{$^1$Materia Kondentsatuaren Fisika Saila, Zientzi 
Fakultatea,
Euskal Herriko Unibertsitatea,\\
644 Posta kutxatila, E-48080 Bilbo, Basque Country, Spain\\
$^2$Donostia International Physics Center (DIPC) and Unidad de F\'\i
sica
Materiales
CSIC-UPV/EHU,\\
Manuel de Lardizabal Pasealekua, E-20018 Donostia, Basque Country,
Spain\\
$^3$Department of Physics, University of Bath, Bath BA2 7AY, United
Kingdom}

\date{\today}

\begin{abstract}
A theoretical many-body analysis is presented of the electron-electron
inelastic lifetimes of Shockley electrons and holes at the (111)
surface of Cu. For a description of the decay of Shockley states both
below and above the Fermi level, single-particle wave functions have
been obtained by solving the Schr\"odinger equation with the use of
an approximate one-dimensional pseudopotential fitted to reproduce
the correct bulk energy bands and surface-state dispersion. A
comparison with previous calculations and experiment indicates that
inelastic lifetimes are very sensitive to the actual shape of the
surface-state single-particle orbitals beyond the $\bar\Gamma$
(${\bf k}_\parallel=0$) point, which controls the coupling between the
Shockley electrons and holes.  
\end{abstract}


\maketitle

\section{Introduction}

A variety of metal surfaces, such as the (111) surfaces of the noble
metals, are known to support a partially occupied band of Shockley
surface states with energies near the Fermi level,\cite{ingles} whose
dynamics have been the subject of long-standing interest.
\cite{mar,eche1,eche2} In particular, the lifetimes of excited holes
at the band edge (${\bf k}_\parallel=0$) of these surface states have
been investigated with high resolution angle-resolved photoemission
(ARP)\cite{arp1,arp2,arp3,arp4} and with the use of the scanning
tunneling microscope (STM).\cite{stm1} STM techniques have also
allowed the determination of the lifetimes of excited Shockley and
image electrons over a range of energies above the Fermi level.
\cite{above1,above2}

\begin{figure}
\includegraphics[width=0.95\linewidth]{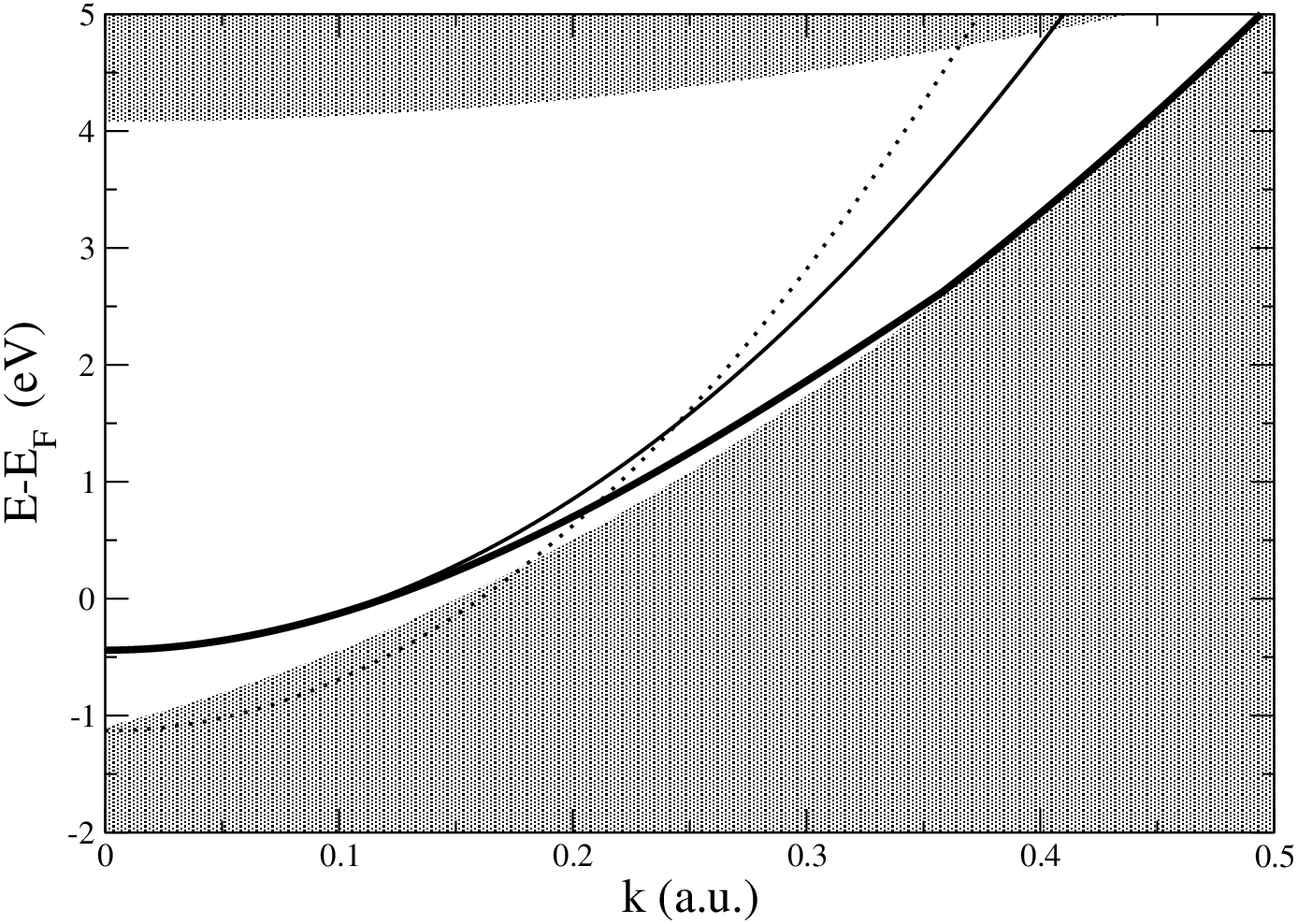}
\caption{Dispersion of the Cu(111) Shockley surface state (thick solid
line), as obtained from three-dimensional {\it ab initio} calculations. Shaded areas represent areas outside the band gap, where bulk
states exist. The thin solid and dotted lines represent approximate
energy dispersions of the Shockley surface state and the bottom of
the projected band gap, respectively, as obtained from
$\varepsilon+k^2/2m$ with $\varepsilon=-0.44\,{\rm eV}$ and $m=0.42$
in the case of the surface state (thin solid line) and
$\varepsilon=-1.09\,{\rm eV}$ and $m=0.22$ for the bottom of the
projected band gap (thin dotted line).}
\label{fig1}
\end{figure}

Many-body calculations of the electron-electron (e-e) inelastic
lifetimes of excited holes at the surface-state band edge of the
(111) surfaces of the noble metals Cu, Ag, and Au, which were based
upon the $G^0W$ one-dimensional scheme that had been introduced by
Chulkov {\it et al.}\cite{chul1} to describe the lifetimes of
image-potential states,\cite{chuli} showed considerable agreement
with experiment.\cite{stm2,chul0} These calculations were then
extended to treat the case of excited surface-state and
surface-resonance electrons above the Fermi level.\cite{chul2,chul3}
In order to account approximately for the potential variation in the
plane of the surface, the original one-dimensional model potential,
which had been introduced to describe surface states at the $\bar\Gamma$ point, was modified along with the
introduction of a realistic effective mass for the dispersion curve
of both bulk and surface states. Within this model, however, all
Shockley states have the same effective mass, so the projected band
structure is not correct, especially at energies above the Fermi
level, as shown in Fig.~\ref{fig1}.  

In this paper, we present a new approach that although at the
$\bar\Gamma$ point is less sophisticated than the model used in
Refs.~\onlinecite{stm2,chul0,chul2,chul3} (i) basically reproduces
the surface-state probability density of
Refs.~\onlinecite{stm2,chul0,chul2,chul3} at the band edge of the
surface-state band in Cu(111) and (ii) has the merit that it reproduces,
through the introduction of a ${\bf k}_\parallel$-dependent
one-dimensional potential, the actual bulk energy bands and
surface-state energy dispersion of Fig.~1, thereby allowing for a realistic
description of the electronic orbitals beyond the $\bar\Gamma$ point.
Adding the contribution from electron-phonon coupling,\cite{eiguren}
which is particularly important at the smallest excitation energies,
our calculations of the lifetime broadening of excited Shockley
electrons and holes in Cu(111) indicate that (i) there is good
agreement with experiment at the surface-state band edge and (ii) at
energies above the Fermi level the lifetime broadening is closer to
experiment and very sensitive to the actual shape of the
surface-state single-particle orbitals beyond the $\bar\Gamma$ point.

Let us consider a semi-infinite many-electron system that is
translationally invariant in the plane of the surface (normal to the
$z$ axis). The decay rate or inverse lifetime of a quasiparticle
(electron or hole) that has been added to the system in the
single-particle state
${\rm e}^{i{\bf k}_i\cdot{\bf r}}\psi_{k_i,E_i}(z)$ of energy $E_i$ is
obtained as follows (we use atomic units, i.e.,
$e^2=\hbar=m_e=1$)\cite{nekovee}
\begin{equation}\label{eq1}
\Gamma_{k_i,E_i}=\mp 2\int dz\int dz'\psi_{k_i,E_i}^{*}(z)\,
{\rm Im}\Sigma(z,z';k_i,E_i)\psi_{k_i,E_i}(z'),
\end{equation}
where the $\mp$ sign in front of the integral should be taken to be
minus or plus depending on whether the quasiparticle is an electron
($E_i>E_F$) or a hole ($E_i\le E_F$), respectively, $E_F$ is
the Fermi energy, ${\bf r}$ and ${\bf k}_i$ represent the position and
wave vectors in the plane of the surface, and $\Sigma(z,z';k_i,E_i)$
is the nonlocal self-energy operator.

To lowest order in a series-expansion of the self-energy
$\Sigma(z,z';k_i,E_i)$ in terms of the energy-dependent screened
interaction
$W(z,z';k,E)$,\cite{hedin,gun} and replacing the interacting Green
function $G(z,z';k,E)$ by its noninteracting counterpart, one finds
the following expression for the imaginary part of the so-called
$G^0W$ self-energy:
\begin{eqnarray}\label{eq2}
&&{\rm Im}\Sigma(z,z';k_i,E_i)={1\over\pi}\int_0^{|E_i-E_F|} dE
\int{d{\bf q}\over(2\pi)^3}\cr\cr
&&\times{\rm Im}G^0(z,z';k_f,E_f)\,{\rm Im}W(z,z';q,E).
\end{eqnarray}
Here, ${\bf k}_f={\bf k}_i-{\bf q}$, $E_f=E_i-E$, and $G^0(z,z';k,E)$
is the noninteracting Green function
\begin{equation}\label{eq3}
G^0(z,z';k,E)=2\,{\psi_{k,E}^+(z^>)\,\psi_{k,E}^-(z^<)\over
\left[\psi_{k,E}^+,\psi_{k,E}^-\right](z)},
\end{equation}
where $z^<$ ($z^>$) is the lesser (greater) of $z$ and $z'$,
and $[f,g](z)$ is the wronskian
\begin{equation}
[f,g](z)=f(z)g'(z)-f'(z)g(z).
\end{equation}
The functions $\psi_{k,E}^\pm(z)$ are solutions of the
single-particle Schr\"odinger equation
\begin{equation}\label{sc}
-(1/2)\,\psi_{k,E}''+V_k(z)\,\psi_{k,E}(z)=
(E-k^2/2)\,\psi_{k,E}(z)
\end{equation} 
regular at $\pm\infty$, with $V_k(z)$ being a momentum-dependent
one-dimensional effective potential that we fit to the projected
surface band structure. We use
\begin{equation}\label{model}
V_k(z)=\cases{U_k+2V_k\cos(2\pi z/a_s),&$z<z_k$\cr\cr
\Phi&$z>z_k$,}
\end{equation} 
where $U_k$ and $V_k$ are fitted to the bulk energy bands (which we have obtained from three-dimensional {\it ab initio} calculations),
$a_s=2.08\,{\rm\AA}$
represents the interlayer spacing, $\Phi=4.94\,{\rm eV}$ is the experimentally
determined work function, and the matching plane $z_k$ is chosen to
give the correct surface-state dispersion represented in Fig.~1 by a
thick solid line.\cite{param} 

In the random-phase approximation (RPA),\cite{fetter} the screened
interaction
$W(z,z';q,E)$ is obtained from the knowledge of the noninteracting
density-response function $\chi^0(z,z';q,E)$ by solving the following
integral
equation
\begin{eqnarray}\label{eqlast}
&&W(z,z';q,E)=v(z,z';q)+\int dz_1\int dz_2\cr\cr
&&\times v(z,z_1;q)\,\chi^0(z_1,z_2;q,E)\,W(z_2,z';q,E),
\end{eqnarray}
where $v(z,z';q)$ represents the two-dimensional Fourier transform of
the bare Coulomb interaction. The results presented below have been obtained by using in $\chi^0(z,z';q,E)$ the eigenfunctions and eigenvalues of the one-dimensional Hamiltonian of Ref.~\onlinecite{chul1} with all effective
masses set equal to the free-electron mass. We have also used the eigenfunctions and eigenvalues of a single-particle jellium-surface Kohn-Sham Hamiltonian (in the local-density approximation) with $r_s=2.67$,\cite{note} and we have found that the surface-state lifetimes are rather insensitive to whether one or the other choice is employed.

\begin{figure}
\includegraphics[width=0.95\linewidth]{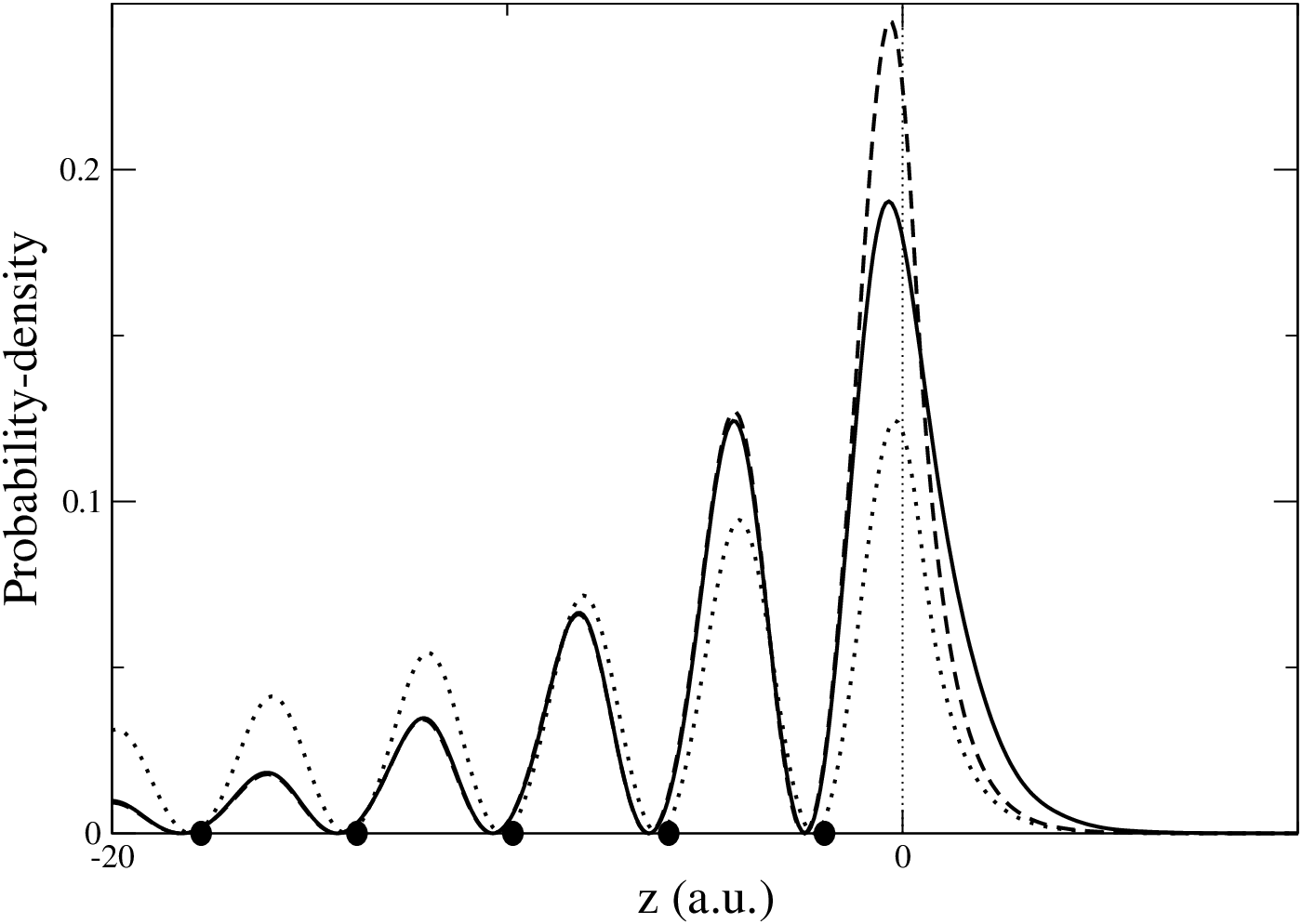}
\caption{Probability-density $|\psi_{k,E}|^2$ of the Shockley
surface state at the center of the surface Brillouin zone of Cu(111),
as obtained within two different models: the model presented here
(dashed line) and the model reported in Ref.~\onlinecite{chul1} (solid line). The dotted line represents the probability-density that we obtain at $k=0.2\,{\rm a.u}$.  Full
circles represent the atomic positions of Cu in the (111) direction.
The geometrical (jellium) electronic edge ($z=0$) has been chosen to be located
half an interlayer spacing beyond the last atomic layer.}
\label{fig2}
\end{figure}   

The abrupt step model potential of Eq.~(\ref{model}), which does not
account for the image tail outside the surface, could not possibly be
used to describe image states. However, Shockley surface states are
known to be rather insensitive to the actual shape of the potential
outside the surface; indeed, the model potential of Eq.~(\ref{model})
is found to yield a surface-state probability-density
$|\psi_{k_i,E_i}|^2$ at the band edge of the Shockley surface-state
band of Cu(111) ($k_i=0$) that is in reasonably good agreement with
the more realistic surface-state probability density used in
Refs.~\onlinecite{stm2,chul0,chul2,chul3}, as shown in Fig.~2. Both
probability-densities coincide within the bulk, although our
approximate probability-density appears to be slightly more localized
near the surface, as expected.\cite{notenew} Nevertheless, we find that decay rates
of an excited hole at $\bar\Gamma$ based on the use of these two
models to describe the wave function $\psi_{k_i,E_i}$ entering
Eq.~(\ref{eq1}) agree within less than $1\,{\rm meV}$. Differences
between our new calculations, which are based on the use of the
$k$-dependent model potential of Eq.~(\ref{model}), and those
reported previously,\cite{stm2,chul0,chul2,chul3} are due to our more
realistic description of the band structure and surface-state wave
functions beyond the $\bar\Gamma$ point.  

\begin{table}
\caption{$G^0W$ decay rates, in linewidth units (meV), of an excited
hole at the band edge of the Shockley surface-state band of Cu(111)
($E_i=-0.44\,{\rm eV}$ and $k_i=0$). $\Gamma_{\rm inter}$ and
$\Gamma_{\rm intra}$ represent interband and intraband contributions
to the e-e decay rate $\Gamma_{\rm e-e}$. The total decay rate
$\Gamma_{\rm total}$ includes the e-ph decay rate of 7 meV reported
in Ref.~\onlinecite{eiguren}. The full calculation represents the
result we have obtained by using in Eq.~(\ref{eq3}) the actual
$k$-dependent model single particle wave function $\psi_{k,E}$. The
approximate calculation represents the result we have obtained by
replacing all surface-state wave functions with ${\bf k}\neq 0$ by
that at the $\bar\Gamma$ point. The third row represents the
calculations reported in Refs.~\onlinecite{stm2,chul0}. The
experimental linewidth has been taken from the STM measurements
reported in Ref.~\onlinecite{stm2}.}
\begin{ruledtabular} \begin{tabular}{lccccc}
Calculation&$\Gamma_{\rm inter}$&$\Gamma_{\rm intra}$ &$\Gamma_{\rm e-e}$&
$\Gamma_{\rm total}$\\ \hline
Full&10&9&19&26\\
Approximate&10&23&33&40\\
Refs.~\onlinecite{stm2,chul0}&6&19&25&32\\
Experiment&&&&24\\
\end{tabular}
\end{ruledtabular} \label{table1}
\end{table}

First of all, we consider an excited hole at the band edge of the
Shockley surface-state band of Cu(111), i.e., with $E_i=-0.44\,{\rm
eV}$ and $k_i=0$ (see Fig.~1). The decay of this excited
quasiparticle may proceed either through the coupling with bulk
states (interband contribution) or through the coupling, within the
surface-state band itself, with surface states of different wave
vector ${\bf k}$ parallel to the surface (intraband contribution). In
order to investigate the impact of the actual shape of the
surface-state wave functions with ${\bf k}\neq 0$ on the decay of the
surface-state hole at $\bar\Gamma$, we have compared in Table I the
decay rates that we have calculated either by using in
Eq.~(\ref{eq3}) the actual $k$-dependent model wave function
$\psi_{k,E}$ (full calculation) or by replacing all surface-state
wave functions $\psi_{k,E}$ with ${\bf k}\neq 0$ by that at the
$\bar\Gamma$ point (approximate calculation). This comparison shows
that the penetration of the actual $k$-dependent surface-state wave
functions $\psi_{k,E}$ being larger than at $\bar\Gamma$ (compare the dashed and dotted lines of Fig.~\ref{fig1}) yields a
reduction in the decay rate from $33\,{\rm meV}$ to $19\,{\rm meV}$, which is due to the fact that the coupling of the surface-state hole at $\bar\Gamma$
with actual surface states of different wave vector ${\bf k}$ (intraband contribution) is smaller than the coupling that would take place with surface-state orbitals that do not change with ${\bf k}$ .
The difference between our predicted surface-state lifetime
broadening of $19\,{\rm meV}$ and that reported
before ($\tau^{-1}=25\,{\rm meV}$)\cite{stm2,chul0} is entirely due to
our more accurate description of (i) the projected band structure and
(ii) the wave-vector dependence of the surface-state wave functions
$\psi_{k,E}(z)$ entering the evaluation of the Green function of
Eq.~(\ref{eq3}).

We have also carried out a full calculation of the decay of an excited hole at $\bar\Gamma$ but replacing the actual surface-state wave vector ${\bf k}_f$ entering Eq.~(\ref{eq2}) by the wave vector that would correspond to a parabolic surface-state dispersion of the form dictated by the thin solid line of Fig.~\ref{fig1}, and we have found that the linewidth is {\it reduced} (as expected, since the parabolic dispersion results in fewer final states) by less than $\sim 1\,{\rm meV}$. However, if one further replaces our wave-vector dependent surface-state orbitals entering Eq.~(\ref{eq2}) by their less accurate counterparts used previously,\cite{stm2,chul0} the lifetime broadening is increased considerably (from $19$ to $25\,{\rm meV}$), showing the important role that the actual coupling between initial and final states plays in the surface-state decay mechanism.

Our model, which correctly reproduces the behaviour of $s$-$p$ valence
states, does not account for the presence of $d$ electrons with
energies a few electronvolts below the Fermi level. The screening of
$d$ electrons is known to play a crucial role in the decay mechanism
of bulk states.\cite{igor} However, in the case of Shockley
holes, whose decay is dominated by intraband transitions that are
associated with very small values of the momentum transfer, the
screening of $d$ electrons is expected to reduce the lifetime broadening
only very slightly,\cite{aran} and will not be included in the present work.
Adding to our  estimated e-e linewidth of $19\,{\rm meV}$ the
electron-phonon (e-ph) linewidth of $7\,{\rm eV}$ reported in Ref.~\onlinecite{eiguren}, we find $\Gamma_{\rm total}=26\,{\rm meV}$ in close agreement with the
experimentally measured linewidth of $24\,{\rm meV}$, as shown in
Table I.

\begin{figure}
\includegraphics[width=0.95\linewidth]{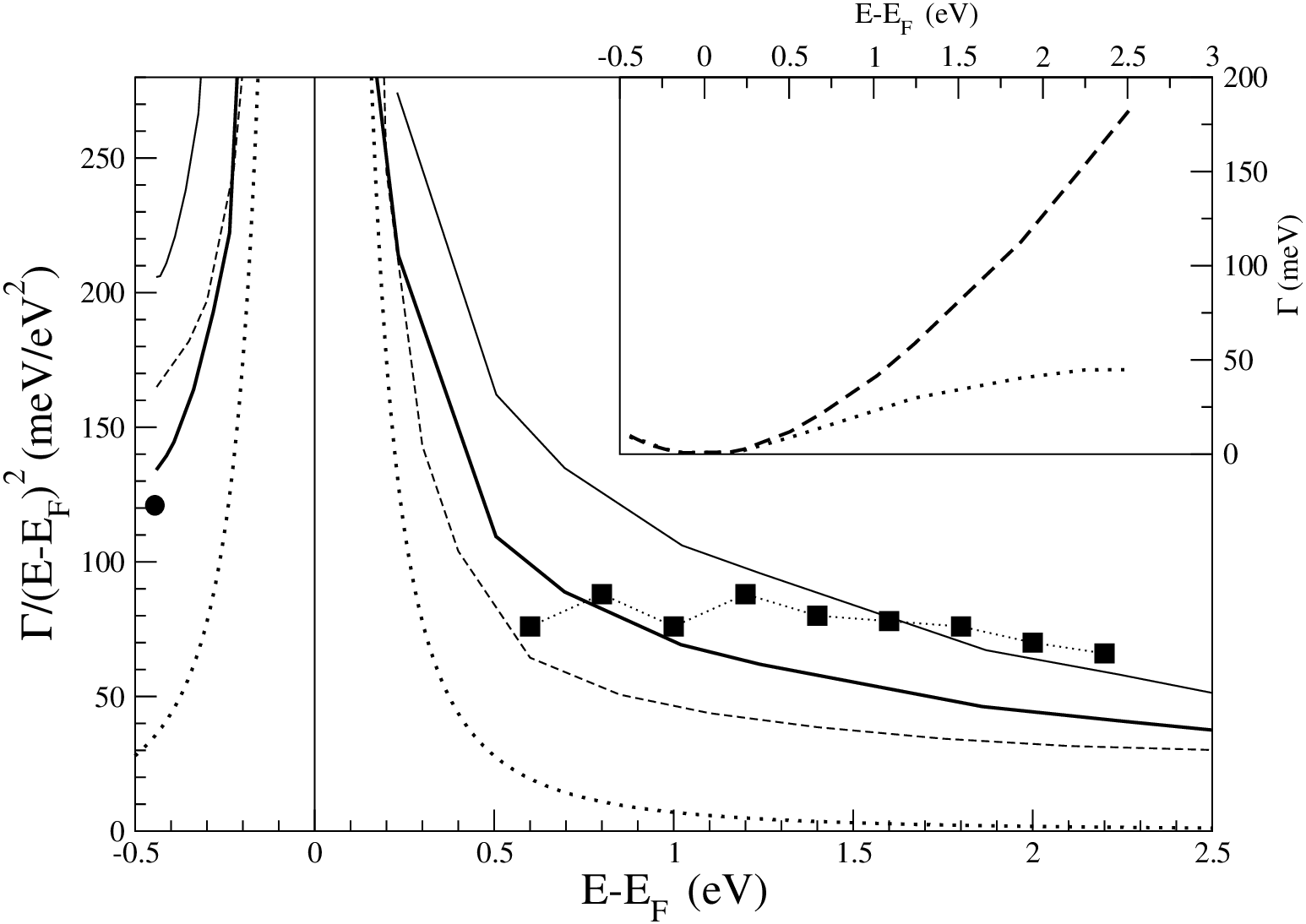}
\caption{Scaled total linewidth $\Gamma_{\rm total}/(E_i-E_F)^2$ of
Shockley surface-state electrons and holes in Cu(111) (including the
e-ph contribution $\Gamma_{\rm e-ph}=7\,{\rm meV}$) versus the surface-state energy $E_i$, as obtained from Eqs.~(\ref{eq1})-(\ref{eqlast}) (thick solid line), from Eqs.~(\ref{eq1})-(\ref{eqlast}) but replacing all surface-state wave functions with ${\bf k}\neq 0$ by those at the $\bar\Gamma$ point (thin solid line), and from Ref.~\onlinecite{chul2} (thin dashed line). The dotted line represents the energy-independent e-ph contribution $\Gamma_{\rm e-ph}=7\,{\rm meV}$. The dotted line with squares represents the STM measurements reported in Ref.~\onlinecite{above1}, multiplied by a
factor of two to correct for an error in the phase coherence length
used in that work. The solid circle represents the experimentally
determined linewidth of an excited hole at the $\bar\Gamma$ point, as
reported in Ref.~\onlinecite{stm2}. Separate unscaled interband
($\Gamma_{\rm inter}$) and intraband ($\Gamma_{\rm intra}$) contributions to
the total linewidth
($\Gamma_{\rm total}=\Gamma_{\rm intra}+\Gamma_{\rm inter}+\Gamma_{\rm e-ph}$) are
represented in the inset by dashed and dottel lines, respectively.}
\label{fig3}
\end{figure}   

In Fig.~3, we show our full calculation (thick solid line) of the
inelastic linewidth ($\Gamma_{\rm total}=\Gamma_{\rm e-e}+\Gamma_{\rm e-ph}$) of
excited Shockley holes and electrons in Cu(111) with energies $E_i$
below and above the Fermi level. Also shown are separate interband and intraband contributions to the
linewidth (inset), the approximate calculation that we have carried out by
replacing all surface-state wave functions with ${\bf k}\neq 0$ by
the surface-state wave function at $\bar\Gamma$ (thin solid line),
and the calculations reported in Ref.~\onlinecite{chul2} (thin dashed
line). The lifetime broadening of excited Shockley electrons in
Cu(111) was studied with the STM by B\"urgi {\it et al.}.
\cite{above1} However, it has been argued recently
\cite{crampin} that due to an error in the identification of the phase
coherence length in the measured quantum interference patterns the
linewidths reported in Ref.~\onlinecite{above1} should be doubled.
These doubled values have been represented in Fig.~3 by solid
squares. The experimentally determined inelastic linewidth of an
excited Shockley hole at the $\bar\Gamma$
point\cite{stm2} is represented by a solid circle.

Fig.~3 shows that a correct description of the wave-vector dependent
surface-state wave functions reduces the coupling of holes and
electrons within the Shockley band, thereby bringing the lifetime
broadening into closer agreement with experiment. A comparison
between our full calculations and experiment shows that there is
close agreement at the surface-state band edge (at $E-E_F=-0.44\,{\rm
eV}$) and there is also reasonable agreement at energies above the
Fermi level. At energies where the surface-state band merges into the
continuum of bulk states, however, our calculated linewidths are
still too low, which might be a signature of the need of a fully
three-dimensional description of the surface band structure. We also
note that differences between the calculations reported here and
those reported previously\cite{chul2} indicate that inelastic lifetimes
are very sensitive to the actual shape of the surface-state
single-particle orbitals beyond the $\bar\Gamma$ point. The linewidths reported here are smaller (larger) for excited holes (electrons) below (above) the Fermi level, thus bringing the theretical predictions closer to experiment.

In summary, we have presented a new $G^0W$ one-dimensional scheme to
calculate the inelastic lifetime broadening of excited Shockley electrons and holes in Cu(111), which is based on a realistic description
of the projected bulk energy bands and the surface-state orbitals
beyond the $\bar\Gamma$ point. Adding the contribution from
electron-phonon coupling,\cite{eiguren} which is particularly
important at the smallest excitation energies, our calculations indicate that there is reasonable agreement with experiment, especially at low excitation energies. The screening of $d$ electrons, not included in this work, is expected to reduce the lifetime broadening only very slightly, at least at the hot-electron energies nearest to the Fermi level. 

The authors thank E. V. Chulkov, P. M. Echenique, and J. E. Inglesfield for enjoyable discussions and a careful reading of the manuscript. M.G.V. and
J.M.P. acknowledge partial support by the University of the Basque
Country, the Basque Unibertsitate eta Ikerketa Saila, the MCyT, and
the EC 6th framework Network of Excellence NANOQUANTA
(NMP4-CT-2004-500198).


\begin{thebibliography}{99}
\bibitem{param}  At the $\bar\Gamma$ point ($k=0$), one finds
$U_k=-7.00\,{\rm eV}$ (with respect to the Fermi level), $2V_k=5.17\,{\rm eV}$, and $z_k=0.18\,{\rm\AA}$ (with respect to the jellium edge). At $k=0.5\,{\rm a.u.}$, we find
$U_k=-4.24\,{\rm eV}$, $2V_k=0.19\,{\rm eV}$, and $z_k=0.37\,{\rm\AA}$.
\bibitem{note} The valence electron density $n$ of the solid is
usually
characterized by the density parameter $r_s=(3/4\pi n)^{1/3}/a_0$,
$a_0$
being the Bohr radius $a_0=0.529\,{\rm\AA}$.
\bibitem{notenew} For the overlap integral we have found
$<\psi_1|\psi_2>=0.99$, $\psi_1$ and $\psi_2$ being the Shockley probability-amplitudes leading to the probability-densities represented in Fig.~\ref{fig2} by solid and dashed lines.
\end{thebibliography}
\end{document}